\documentclass{JINST}
\usepackage{graphicx}

\title{Registration of reactor neutrinos with the highly segmented plastic scintillator detector DANSSino}

\author{
V.~Belov$^a$,
V.~Brudanin$^a$,
M.~Danilov$^{b,c}$,
V.~Egorov$^a$\thanks{Corresponding author.},
M.~Fomina$^a$,
A.~Kobyakin$^b$,
V.~Rusinov$^b$,
M.~Shirchenko$^a$,
Yu.~Shitov$^a$,
A.~Starostin$^b$,
I.~Zhitnikov$^a$\\
\llap{$^a$}Joint Institute for Nuclear Research, Dubna 141980, Russia\\
\llap{$^b$}State Scientific Center, Institute for
 Theoretical and Experimental Physics, Moscow, Russia\\
\llap{$^c$}Moscow Institute of Physics and Technology, Moscow Region 141700, Russia\\
  E-mail: \email{egorov@jinr.ru}}
\abstract{DANSSino is a simplified pilot version of a solid-state detector of reactor antineutrino (it is being created within the DANSS project and will be installed close to an industrial nuclear power reactor). Numerous tests performed under a 3~GW$_{\rm th}$ reactor of the Kalinin NPP at a distance of 11 m from the core demonstrate operability of the chosen design and reveal the main sources of the background. In spite of its small size ($20\times20\times100$~cm$^3$), the pilot detector turned out to be quite sensitive to reactor neutrinos, detecting about 70 IBD events per day with the signal-to-background ratio about unity.}

\begin{document}
\section{Introduction}
\begin{figure}[th]
 \setlength{\unitlength}{1mm}
 \begin{picture}(150,60.0)(0,0)
  \put(8,5){\includegraphics{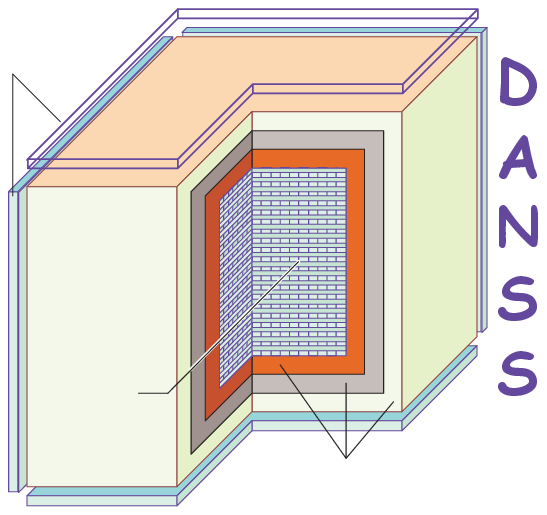}}
  \put(10,20){\parbox{15mm}{\begin{center}\tiny\sf Segmental polystyrene-based solid plastic scintillator\\[1mm]
  1~m$^3$\\2500 strips \end{center}}}
  \put(43,9.5){\makebox(0,0)[t]{\scriptsize\sf Cu+Pb+CHB}}
  \put(43,6.5){\makebox(0,0)[t]{\tiny\sf passive shielding}}
  \put(7,52){\parbox{15mm}{\scriptsize\sf Muon veto\\plates}}
  \put(100.0,0){\includegraphics{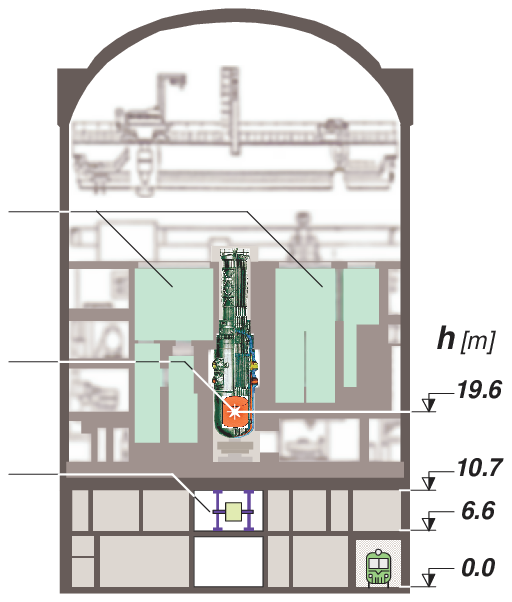}}
  \put(0.9,2.2){\includegraphics{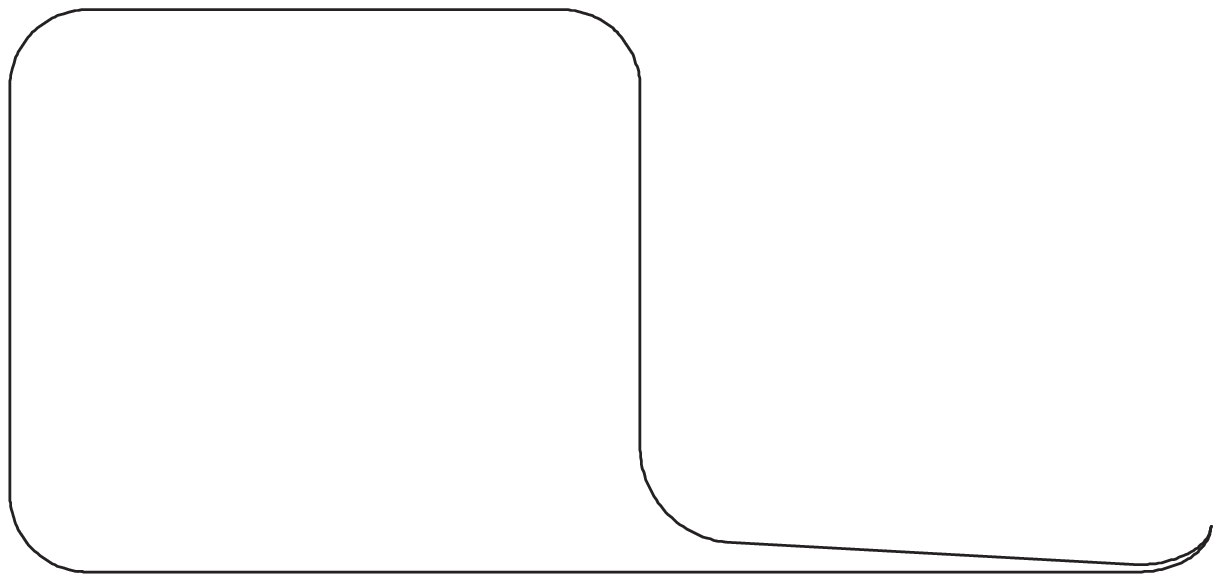}}
    \put(72,53.0){\parbox{40mm}{\begin{center}\footnotesize\sf A typical WWER-1000 \\[-0.5mm]reactor building\end{center}}}
    \put(75,36.7){\parbox{30mm}{\begin{center}\scriptsize\sf Reservoirs with\\[-0.5mm] technological liquids\end{center}}}
    \put(72,23.7){\parbox{30mm}{\begin{center}\scriptsize\sf A core of the reactor:\\[-0.5mm] $\varnothing$ 3.12 m $\times$ h 3.55 m\end{center}}}
    \put(73,12.5){\parbox{30mm}{\begin{center}\scriptsize\sf A movable platform\\[-0.5mm] with a lifting gear\\[-0.5mm] in a service room\end{center}}}
 \end{picture}
 \caption{Position of the DANSS neutrino detector under an industrial reactor WWER-1000.}
 \label{Fig.DANSS}
 \end{figure}

The DANSS project \cite{DANSS_AAP,DANSS_TAUP,DANSS_ICHEP} proposes to create a relatively compact neutrino spectrometer which does not contain any flammable or other dangerous liquids and may therefore be located very close to a core of an industrial power reactor (Fig.~\ref{Fig.DANSS}). Due to a high neutrino flux ($\sim5\times10^{13}\; \bar\nu_e /{\rm cm}^2/{\rm s}$ at a distance of 11 m) it could be used for the reactor monitoring and for fundamental research including neutrino oscillation studies.

 \begin{figure}[th]
 \setlength{\unitlength}{1mm}
 \begin{picture}(150,53)(0,0)
  \put(24,2){\includegraphics{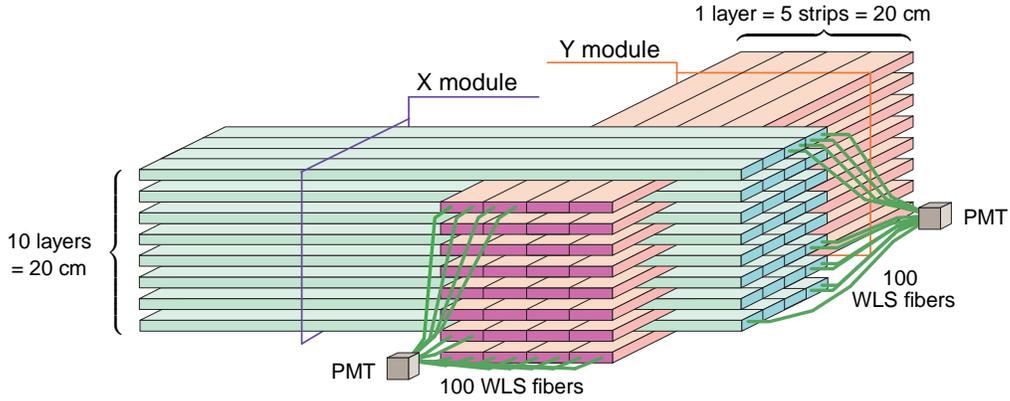}} 
  \put(68.0,42.0){\makebox(0,0)[c]{\footnotesize\sf X module}}
  \put(87.0,46.4){\makebox(0,0)[c]{\footnotesize\sf Y module}}
  \put(5.0,16.0){\parbox{15mm}{\begin{center}\scriptsize\sf 10 layers\\ = 20 cm\end{center}}}
  \put(23.0,19.5){\makebox(0,0.5)[r]{\scriptsize\sf $\left\{\rule{0mm}{12.5mm}\right.$}}
  \put(53.0,2.5){\makebox(0,0)[b]{\scriptsize\sf PMT}}
  \put(74.0,0.5){\makebox(0,0.5)[b]{\scriptsize\sf 100 WLS fibers}}
  \put(140.0,24.0){\makebox(0,0)[r]{\scriptsize\sf PMT}}
  \put(128.0,16.0){\makebox(0,0)[r]{\scriptsize\sf 100}}
  \put(133.0,13.0){\makebox(0,0)[r]{\scriptsize\sf WLS fibers}}
  \put(114.0,49.5){\makebox(0,0)[b]{\scriptsize\sf 1 layer = 5 strips = 20 cm}}
  \put(115.5,47.0){\makebox(0,0)[b]{\scriptsize\sf $\overbrace{\rule{23mm}{0mm}}$}}
 \end{picture}
 \caption{Two of fifty intercrossing modules of the DANSS neutrino detector.}\label{Fig.DANSS_Modules}
 \end{figure}

The DANSS detector will consist of highly segmented plastic scintillator with a total volume of 1 m$^3$, surrounded with a composite shield of lead, copper and borated polyethylene, and vetoed against cosmic muons with a number of external scintillator plates.
The basic element of DANSS is a polystyrene-based extruded scintillator strip ($1\times4\times100$~ cm$^3$) with a thin Gd-containing surface coating which is a light reflector and an ($n,\gamma$)-converter simultaneously.
The coating (about 0.1--0.2~mm) is produced by co-extrusion and consists of polystyrene with 18\% admixture of TiO$_2$ and 6\% of Gd$_2$O$_3$, so that final Gd density is about 1.6~mg/cm$^2$ $\sim$ 0.35\%$_{\rm wt}$. Light collection from the strip is done via three wavelength-shifting Kuraray fibers Y-11, $\varnothing$~1.2~mm, glued into grooves along all the strip. An opposite (blend) end of each fiber is polished and covered with a mirror paint, which decreases a total lengthwise attenuation of a light signal down to $\sim$5~\%/m.

Each 50 parallel strips are combined into a module, so that the whole detector (2500 strips) is a structure of 50 intercrossing modules (Fig.~\ref{Fig.DANSS_Modules}). Each module is viewed by a compact photomultiplier tube (Hamamatsu R7600U-200) coupled to all 50 strips of the module via 100 WLS fibers, two per a strip. In addition, to get more precise energy and space pattern of an event, each strip is equipped with an individual multipixel photosensor (SiPM) operating in the Geiger mode and coupled to the third WLS fiber.

In order to check operability of the DANSS design, compare different acquisition schemes, and measure the real background conditions, a simplified pilot version of the detector (DANSSino) was created. Below we present the DANSSino description and some results of numerous tests performed at the JINR laboratory and under the KNPP reactor.

\section{The DANSSino design}
\begin{figure}[ht]
 \setlength{\unitlength}{1mm}
 \begin{picture}(150,52)(0,0)
  \put(1,0){\includegraphics{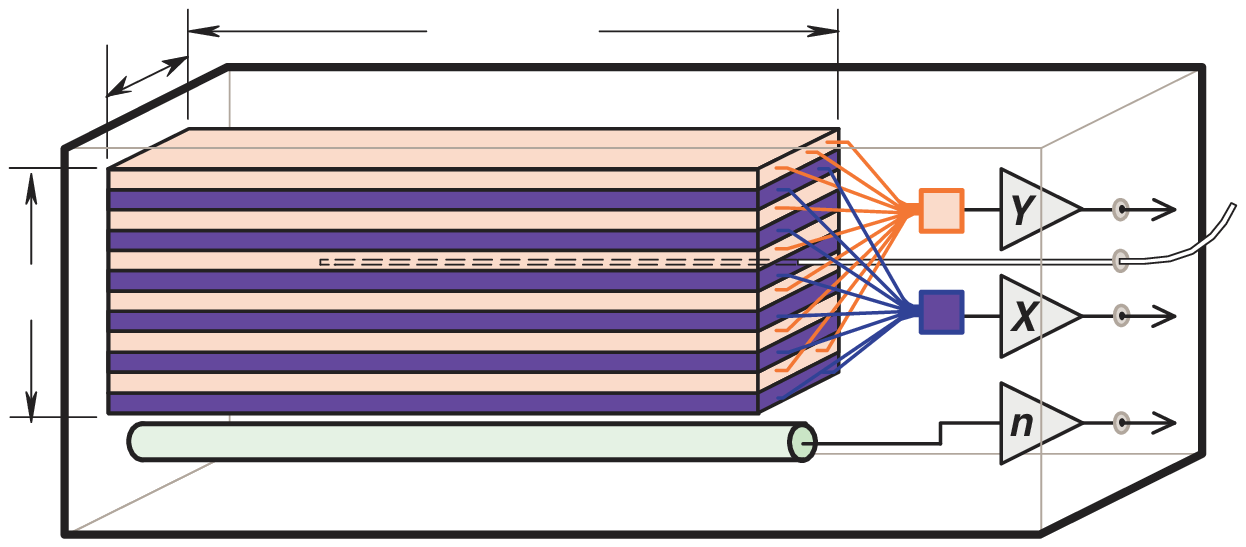}}   
  \put(125.0,3){\includegraphics{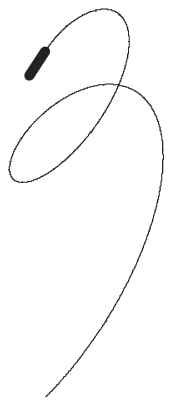}}  
  \put(127.5,47.0){\parbox{20mm}{\scriptsize\sf Calibration\\source}}
  \put(52.5,52.0){\makebox(0,0)[c]{\scriptsize\sf 100 cm}}
  \put(3.5,25.5){\makebox(0,0)[c]{\scriptsize\sf 20}}
  \put(15.0,49.5){\makebox(0,0)[c]{\scriptsize\sf 20}}
  \put(96.0,37.7){\makebox(0,0)[c]{\scriptsize\sf Y\hspace{0.2mm}-PMT}}
  \put(96.0,18.8){\makebox(0,0)[c]{\scriptsize\sf X-PMT}}
  \put(52.0,4.0){\makebox(0,0)[b]{\scriptsize\sf $^3$He gas-based neutron counter}}
 \end{picture}
 \caption{The DANSSino detector}\label{Fig.DANSSino}
 \end{figure}

DANSSino consists of exactly the same basic elements as the main DANSS detector.
One hundred strips of DANSSino (Fig.~\ref{Fig.DANSSino}) form a $20\times20\times100$ cm$^3$ bar divided into two modules: the odd layers are coupled to the X-PMT and the even ones to the Y-PMT. Together with an additional neutron counter\footnote{This $^3$He gas-based counter allows high sensitivity monitoring of the thermal neutron flux inside the detector shielding.} both modules are equipped with simple front-end preamplifiers and placed into a light-tight box. Information from the individual photodiodes of each strip is not used in this prototype. To perform energy calibration, a teflon tube is placed inside the bundle of strips, so that a tiny radioactive source can be inserted in the detector with a thin flexible string. For this purpose several gamma and neutron sources with activity of few Bq were produced: $^{137}$Cs, $^{60}$Co, $^{22}$Na, $^{248}$Cm. To suppress external background caused by gamma-rays and thermal neutrons the detector is surrounded with a passive shield. As it is rather compact, the composition of the shielding can be easily changed. Four big scintillator plates ($200\times50\times3$ cm$^3$) similar to that employed in the GERDA experiment \cite{GERDA} are used to tag the events associated with cosmic muons.

The Inverse Beta-Decay (IBD) of hydrogen atoms in the detector body is used to detect the reactor antineutrino:
\begin{equation}
 \widetilde{\nu_e}+p\rightarrow e^++n\;.
\end{equation}
The detection process proceeds in two steps: the first one applies to the positron and the second to the neutron. The energy threshold of the IBD reaction is 1.8 MeV, while most of the remaining neutrino energy is transferred to the positron. The positron deposits its energy within a short range of few cm and then annihilates emitting two 511~keV photons at 180$^\circ$. As a result, the first (Prompt) energy deposit is distributed in space in a very specific way. The second (Delayed) step is the detection of the neutron. Initial energy of the neutron is only few keV. After moderation in the plastic scintillator it is captured by $^{157}$Gd or $^{155}$Gd with a very high cross-section. In both cases a cascade of $\gamma$-rays is emitted with the total energy of about 8 MeV. Because of high multiplicity and deep penetration in plastic these $\gamma$-rays produce a flash which is spread widely within a sphere with a diameter of about 30-40 cm, so that a number of strips in both X and Y modules are usually fired. Distribution of time between the Prompt and Delayed signals depends on the detector structure and is described by a combination of two exponents with the characteristic times $\tau_m$ and $\tau_c$ that correspond to the neutron moderation and capture respectively.

Simple acquisition system allows discrimination of events consisting of (Prompt + Delayed) signal pairs. The energy of both Prompt and Delayed signals ($E_P$ and $E_D$) detected by both X and Y modules are measured separately with two 2-fold QDCs. Finally, each collected event contains 4 energies ($E_{XP}$, $E_{YP}$, $E_{XD}$, $E_{YD}$), time between the P and D pulses ($T_{PD}$), and information about the muon veto (which of the veto plates were fired and when).

\section{Measurements}
\subsection{Above-ground tests at the JINR laboratory}
Numerous tests with the DANSSino detector were performed at the ground floor of the JINR laboratory building (Dubna). Measurements with $^{137}$Cs, $^{22}$Na, and $^{60}$Co $\gamma$-sources compared with the MC simulations were used to calibrate the energy scale and to set the lower energy thresholds of the discriminators.
As the light collection from a long strip is far from being perfect, the number of photoelectrons (p.e.) produced at the PMT photocathode is relatively small. Therefore, the energy resolution of the spectrometer is determined mainly by the Poisson dispersion of this number and depends on the energy deposit in the scintillator which corresponds to a single photoelectron $\varepsilon_{\rm pe}$. In order to estimate it, single-photoelectron noise peaks were measured with both X and Y modules. As a result, $\varepsilon_{\rm pe}$ was found to be about 100 keV/p.e.

\begin{figure}[hb]
 \setlength{\unitlength}{1mm}
 \begin{picture}(150,45)(0,0)
  \put(10,0){\includegraphics{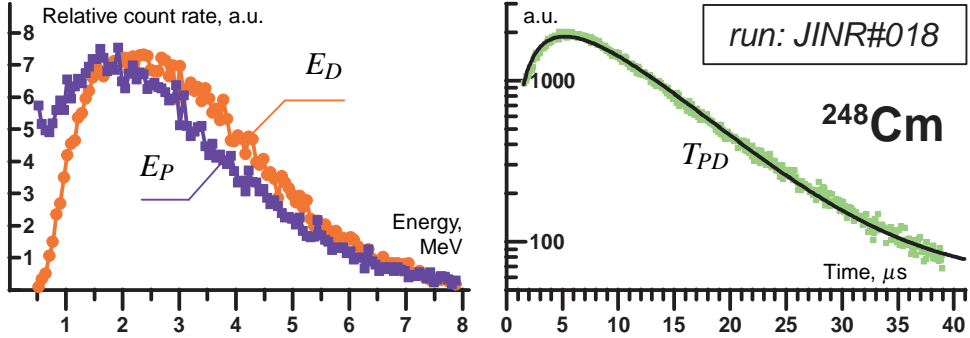}} 
  \put(15,42.5){\makebox(0,0)[l]{\scriptsize\sf Relative count rate, a.u.}}
  \put(71.0,15.0){\makebox(0,0)[r]{\scriptsize\sf Energy,}}
  \put(71.0,12.0){\makebox(0,0)[r]{\scriptsize\sf MeV}}
  \put(50.0,36.0){\makebox(0,0)[l]{\large $E_D$}}
  \put(28.0,22.5){\makebox(0,0)[l]{\large $E_P$}}
  \put(79.0,42.5){\makebox(0,0)[l]{\scriptsize\sf a.u.}}
  \put(130.0,7.0){\makebox(0,0)[rb]{\scriptsize\sf Time, $\mu$s}}
  \put(100.0,24.0){\makebox(0,0)[l]{\large $T_{PD}$}}
 \end{picture}
 \caption{Energy spectra (left) and time distribution (right) of the neutrino-like events measured with the $^{248}$Cm source.  }\label{Fig.ET(248Cm)}
 \end{figure}

The next important parameter of the spectrometer is its selective sensitivity to neutrons. The $^{248}$Cm neutron source provides signals with a signature very similar to the neutrino-like events. Indeed, the Delayed signal is caused by the same $^{157,155}$Gd($n,\gamma$) reaction, whereas the Prompt signal corresponds either to the ($n$-$p$) scattering of the initial fast neutron on hydrogen in the plastic detector or to the prompt $\gamma$-rays following the $^{248}$Cm fission. Therefore, the shape of the Delayed energy spectrum (Fig.~\ref{Fig.ET(248Cm)}) is exactly the same as expected for the IBD process whereas the Prompt spectrum is somewhat different.

The $T_{PD}$ time distribution is also very similar to the IBD. The difference is due to much higher neutron multiplicity ({\sl k}$\simeq$3), so that it is described by a more complicated function, which reflects the fact that only the first of $k$ neutrons can be detected in our case.
Analysis of the $^{248}$Cm data shown in Fig.~\ref{Fig.ET(248Cm)} with $k$=3 gives the following values for the neutron moderation and capture specific times:
\begin{equation} \label{Eq.tau(Cm)}
\tau_m\simeq3\pm1\;{\rm \mu s}\hspace{20mm}
\tau_c\simeq24\pm1\;{\rm \mu s}
\end{equation}

Measurement without radioactive sources shows that in addition to $\gamma$-rays, muons and thermal neutrons, the laboratory natural background contains a significant number of neutrino-like events ($\simeq300$ per hour) consisting of both Prompt and Delayed signals -- (P+D)-pairs. The $E_D$ energy spectrum and $T_{PD}$ time distribution of those events are of the same shape as in the $^{248}$Cm case. It is natural to assume that the events are caused by fast neutrons which could not be rejected even with 16 cm of borated polyethylene. Indeed, it is known \cite{Heusser,AtmosphereBG} that the high-energy hadronic component of the cosmic background at sea level could be as high as $10^2$/m$^2$/s depending on the energy and the site, and a low-Z shielding equivalent to at least 20 meters of water (20 m w. e.) is needed to suppress it.

\subsection{Tests under the KNPP industrial reactor}
After the laboratory tests the DANSSino detector was transported to the Kalinin Nuclear Power Plant (Udomlya, 285 km NW of Dubna) and installed in a service room under Unit\#3 at a distance of 11 m from the core center (Fig.~\ref{Fig.DANSS}). Unit\#3 includes a typical Russian industrial water-moderated water-cooled power reactor WWER-1000 of thermal power 3~GW. Huge water reservoirs with technological liquids, thick walls of heavy concrete, the reactor body and equipment placed above the room provide excellent shielding ($\simeq$50 m w. e.) which completely removes cosmic fast neutrons. The muon component is suppressed by a factor of $\simeq$6. The gamma background at the detector site is slightly higher than at the JINR laboratory, mainly because of high $^{40}$K contamination of the surrounding concrete. Thermal and epithermal reactor neutrons penetrate the room through the monitoring tubes and increase the average background as well. Fortunately, the flux of those neutrons measured with the external four-fold $^3$He gas-based neutron detector turned out to be not too high and could be suppressed with $\sim$10 cm of borated polyethylene (CHB); in addition, slow neutrons cannot emulate the IBD process because of their small energy.

\begin{table}[hb]
\caption{Background in the JINR laboratory and under the KNPP reactor measured by DANSSino unshielded, as well as shielded with 10 cm of Pb, 16 cm of CHB and $\mu$-veto plates (here and below the shielding composition is enumerated from inside to outside).}
\label{Tab.BG}
\begin{center}
\begin{tabular}{|l||r|r||r|r|} \hline
Detector site & \multicolumn{2}{|c||}{ JINR } & \multicolumn{2}{|c|}{ KNPP }\\ \hline
Shielding     &\multicolumn{1}{|c|}{no} &\multicolumn{1}{|c||}{\footnotesize\sf Pb+CHB} &\multicolumn{1}{|c|}{no}&\multicolumn{1}{|c|}{\footnotesize\sf Pb+CHB}\\ \hline \hline
Single signals              &\multicolumn{4}{|c|}{Number of signals per second}\\ \hline
X\hfill{ }{\footnotesize\sf(E$>$0.25 MeV)}   & 532 & 61 & 1470 & 20 \\ \hline              Y\hfill{ }{\footnotesize\sf(E$>$0.25 MeV)}   & 465 & 58 & 1360 & 19 \\ \hline
X$\wedge$Y \hfill{ }{\footnotesize\sf(E$>$0.5 MeV)}   & 235 & 42 &   408 & 11 \\ \hline
X$\wedge$Y \hfill{ }{\footnotesize\sf(E$>$8.0 MeV)}   &  19 & 17 &     4 &  2 \\ \hline \hline
(P+D)-pairs               &\multicolumn{4}{|c|}{Number of (P+D)-pairs per hour}\\ \hline
Total                                 &25058&1657&493230&93\\ \hline
Tagged by $\mu$-veto                 &     & 376&      &52\\ \hline
Free of $\mu$-veto                    &     &1281&      &41\\ \hline
\end{tabular}
\end{center}
\end{table}

The results of some background measurements performed at the JINR laboratory and under the operating KNPP reactor with and without the same shielding are presented\footnote{Some of the JINR and KNPP measurement conditions (the PMT high voltage, discriminator thresholds, energy scale, etc.) could be slightly different, so that the values presented in the Table should be considered as illustrative only.} in Table~\ref{Tab.BG}. From the single count rates of both X and Y modules it is seen that although the initial background conditions under the reactor are worse, a relatively thin passive shielding improves it significantly and makes even 3 times better than in the laboratory. The next two rows represent mainly the flux of thermal neutrons and cosmic muons, respectively (the neutrons produce X$\wedge$Y coincidences with the total energy between the threshold and 8 MeV, whereas the muons also cause X$\wedge$Y coincidences but saturate the QDCs).

Finally, the last two rows in Table~\ref{Tab.BG} show the rate of (P+D) signal pairs without any additional selection. A big number of these events for the unshielded detector arises from random coincidences caused by the high raw count rate. In the case of the shielded detector those are signals mainly from the fast neutrons. As expected, the number of these false neutrino-like events under the reactor becomes much lower than at the laboratory, but still differs from zero. The remaining part is mostly associated with the $\mu$-veto and corresponds probably to secondary fast neutrons produced by cosmic muons in the surrounding heavy materials (Pb and Cu). This assumption was confirmed by measurements performed with different composition of the passive shielding. These data were analyzed under relatively strong requirements which correspond to the IBD signature:
\begin{itemize}
\item the time between the Prompt and Delayed signals must be within a reasonable range ($T_{PD}\in \left[1.5-30.0\right]\,\mu$s);
\item the Delayed signal should correspond to the Gd($n$,$\gamma$) reaction, i.e., both the X and Y detector modules should be fired ($X_D\wedge Y_D$) with a reasonable\footnote{As the detector is small, significant part of the $\gamma$-cascade is not detected, and therefore the acceptable $E_D$ range is extended to the lower energy.} total energy ($E_{XD}+E_{YD}=E_D \in\left[1-8\right]$~MeV);
\item the Prompt signal must also have an appropriate energy: ($E_{XP}+E_{YP} = E_P \in\left[1-7\right]$~MeV).
\end{itemize}

As a result, it was shown that 5 cm of copper in the close vicinity to the detector body without intermediate neutron moderator increase the number of false neutrino-like events associated with muons by a factor of 2, and 10 cm of lead increase it by a factor of 4.

Two of our measurement runs encountered interruptions in the reactor operation (OFF periods), so that it was possible to estimate background experimentally without the neutrino flux\footnote{Actually, when the reactor stops, neutrinos are still emitted by the radioactive fission products \cite{Nu-evolution}, but the decay energy of long-lived nuclei is low and those OFF neutrinos are not detected because of the energy threshold.}.

 \begin{figure}[hb]
 \setlength{\unitlength}{1mm}
 \begin{picture}(150,42)(0,0)
  \put(9,0){\includegraphics{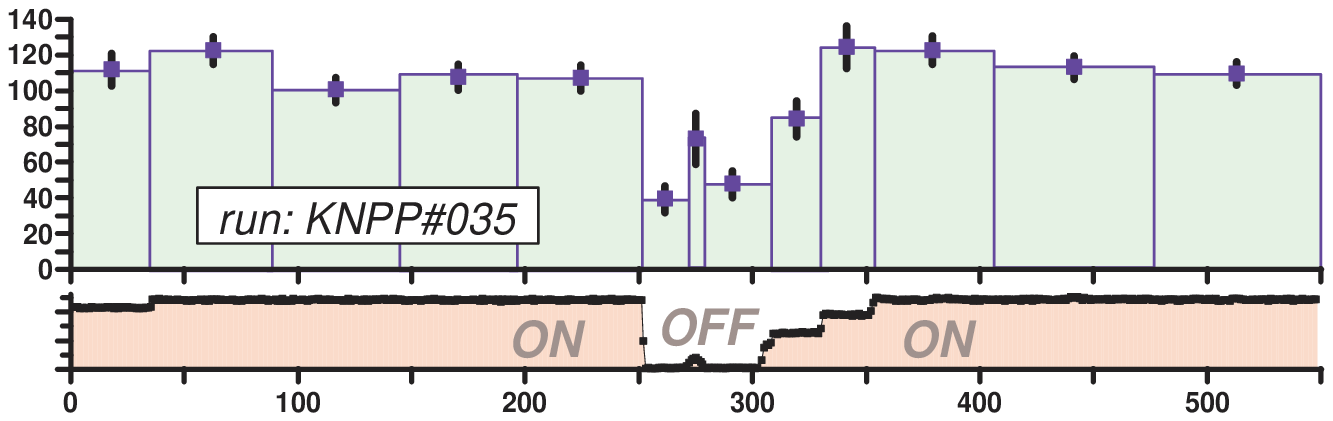}}
  \put( 17.0,42.0){\makebox(0,0)[lt]{\scriptsize\sf Number of events per day}}
  \put( 17.0,8.0){\makebox(0,0)[l]{\scriptsize\sf Relative reactor power}}
  \put(141.0,6.7){\makebox(0,0)[r]{\scriptsize\sf Time,}}
  \put(141.0,1.5){\makebox(0,0)[r]{\scriptsize\sf hr}}
 \end{picture}
 \begin{picture}(150,47)(0,0)
  \put(9,0){\includegraphics{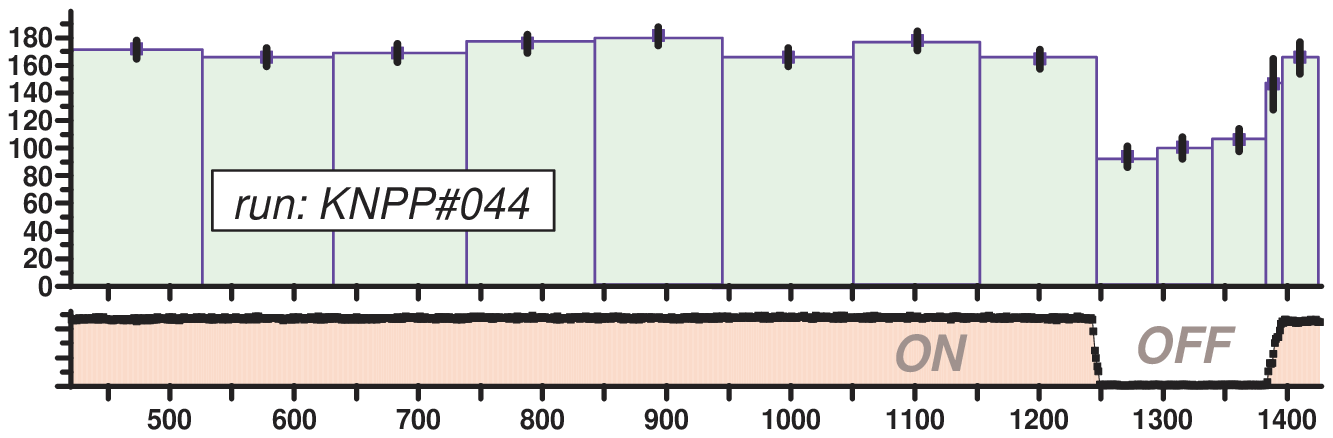}}
  \put(17.0,44.0){\makebox(0,0)[lt]{\scriptsize\sf Number of events per day}}
  \put(17.0,8.0){\makebox(0,0)[l]{\scriptsize\sf Relative reactor power}}
  \put(146.5,6.9){\makebox(0,0)[r]{\scriptsize\sf Time,}}
  \put(146.0,1.8){\makebox(0,0)[r]{\scriptsize\sf hr}}
 \end{picture}
 \caption{Time dependence of the reactor power (bottom of each diagram) and the number of the neutrino-like events detected by DANSSino (top of each diagram) for two measurement periods.}\label{Fig.Nu(time)}
 \end{figure}

Figure~\ref{Fig.Nu(time)} shows time diagrams of these runs. The lower parts of the diagrams show the actual reactor power measured with the external high-sensitive neutron counter installed near the outlet of the monitoring tube. Strong correlation between the reactor power and the number of the neutrino-like events detected with DANSSino is obvious.

Numerical values of these count rates, Signal-to-Background ratios, and muon-induced background rates are given in Table~\ref{Tab.Nu(ON-OFF)}, which describes the neutrino-like events detected with a low energy threshold $E_p^{\rm min}$= 1.5~MeV. It can be seen from the table that intensities of the muon-induced background events do not differ significantly for the ON and OFF periods, which confirms the adequate operation of the acquisition system. A heavy shielding (5 cm of copper) without an inner moderator increases the background by $\sim$70\% but also improves the efficiency\footnote{The efficiency here implies a probability to detect the IBD if it occurs.} by $\sim$10\%, returning part of escaped IBD neutrons back to the scintillator.

\begin{table}[ht]
\caption{Rate of the neutrino-like events in the reactor ON and OFF periods. The shielding thickness (cm) is given in parentheses.}
\label{Tab.Nu(ON-OFF)}
\begin{center}
\begin{tabular}{|l||r|r||r|r|} \hline
Run\# & \multicolumn{2}{|c||}{ KNPP\#035 } & \multicolumn{2}{|c|}{ KNPP\#044 }\\ \hline
Shield composition & \multicolumn{2}{|c||}{\footnotesize\sf CHB(8)+Pb(10)}&
                \multicolumn{2}{|c|}{\footnotesize\sf Cu(5)+CHB(8)+Pb(5)} \\ \hline
Reactor operation &\multicolumn{1}{|c|}{OFF} &\multicolumn{1}{|c||}{ON} &\multicolumn{1}{|c|}{OFF}&\multicolumn{1}{|c|}{ON}\\ \hline \hline
Type of $\nu$-like events &\multicolumn{4}{|c|}{Number of $\nu$-like events per day}\\ \hline
Associated with $\mu$                 & 175$\pm$9&179$\pm$3&318$\pm$8&302$\pm$3\\ \hline
Free of $\mu$-veto                    &  46$\pm$5&108$\pm$3& 94$\pm$4&163$\pm$2\\ \hline \hline
Signal=ON--OFF & \multicolumn{2}{|c||}{62$\pm$5}&\multicolumn{2}{|c|}{70$\pm$5} \\ \hline              S/B=S/OFF      & \multicolumn{2}{|c||}{1.33$\pm$0.25}&\multicolumn{2}{|c|}{0.74$\pm$0.08} \\  \hline
\end{tabular}
\end{center}
\end{table}

\begin{figure}[hb]
 \setlength{\unitlength}{1mm}
 \begin{picture}(150,52)(0,0)
 \put(10,0){\includegraphics{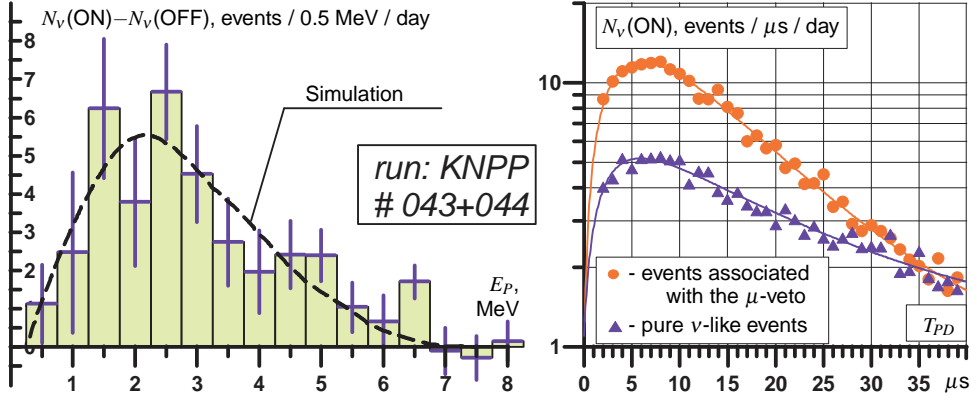}}
 \put(15.0,49.9){\makebox(0,0)[l]{\scriptsize\sf $N_\nu$(ON)$-N_\nu$(OFF), events / 0.5~MeV / day}}
 \put(78.5,14.5){\makebox(0,0)[r]{\scriptsize\sf $E_P$,}}
 \put(78.5,11.5){\makebox(0,0)[r]{\scriptsize\sf MeV}}
 \put(89.5,48.8){\makebox(0,0)[l]{\scriptsize\sf $N_\nu$(ON), events / $\mu$s / day}}
 \put(136.5,9.0){\makebox(0,0)[r]{\scriptsize\sf $T_{PD}$}}
 \put(138.3,0.3){\makebox(0,0)[rb]{\scriptsize\sf $\mu$s}}
 \put(50.0,39.0){\makebox(0,0)[lb]{\scriptsize\sf Simulation}}
 \put(93.0,16.2){\makebox(0,0)[l]{\scriptsize\sf- events associated}}
 \put(98.0,12.9){\makebox(0,0)[l]{\scriptsize\sf with the $\mu$-veto}}
 \put(93.0,09.0){\makebox(0,0)[l]{\scriptsize\sf- pure $\nu$-like events}}
 \end{picture}
 \caption{The differential $E_P$ energy spectrum (left) and $T_{PD}$ time distribution (right) of the neutrino-like events. The dashed curve represents a typical example of the IBD positron energy spectrum \cite{Nu-spectra-old,Nu-spectra-new} calculated for the $^{235}$U fission.}\label{Fig.Ep_Nu(ON-OFF)}
 \end{figure}

Assuming that the OFF data correspond to a pure background, one can build an energy spectrum of IBD positrons as a difference of two: $S$=$N_\nu$(ON)--$N_\nu$(OFF). As an example, $E_P$ and $T_{PD}$ spectra built under an additional ($X_P\!\wedge\! Y_P$) requirement\footnote{Otherwise, the low-energy part of the spectrum becomes polluted with random ($\gamma$-$n$) coincidences.} are shown in Fig.~\ref{Fig.Ep_Nu(ON-OFF)}. In spite of poor statistics and merely illustrative character of the spectra, they are in a very good agreement with theoretical expectations\footnote{For example MC simulations with GEANT-4 predict 75 events/day for the run \#044 while 70$\pm$5 events were observed.}. The steeper slope of the false $T_{PD}$ time curve (upper)  with respect to the true IBD one (lower) is explained by the higher neutron multiplicity ($k\simeq1.6$) of the muon-induced reactions in the copper and lead shielding.

\section{Summary}
In spite of the small size and high edge effects, incomplete shielding and simplified acquisition system, DANSSino is able to detect reactor antineutrinos with the signal-to-background ratio about unity and the efficiency at a level of 10\%, which is in good agreement with the MC simulations.  As the full-scale DANSS detector is of much larger volume, its response function is expected to be considerably better and efficiency significantly higher ($\simeq$70\%) because of a lower relative contribution from the edge detector parts (fewer neutrons and $\gamma$-rays after $n$-capture in Gd would leave the sensitive volume without detection). Together with additional energy and space information from the individual photosensors, it will allow to suppress the background down to a negligible value.

Operation of the detector at a shallow depth with overburden less than 10-20 m w.e. seems to be questionable, because cosmic ray neutrons cannot be tagged by the veto-system. They produce signatures very similar to IBD events but outnumber them by orders of magnitude.

A more detailed description of the detector and the data analysis (including description of the acquisition system and hardware triggers, MC simulations, comparison of the background conditions and different shielding composition) is being prepared and will be published soon.

\section{Acknowledgments}
The work was supported in part by the JINR (grant 13-202-05), Russian Foundation for Basic Research (grants 11-02-01251 and 11-02-12194), the Russian Ministry of Education and Science (grants 8174, 8411, 1366.2012.2).
The authors are grateful to the staff of the Kalinin Nuclear Power Plant, and especially to the KNPP Directorate for providing a possibility of performing research measurements extremely close to the reactor core and to the KNPP Radiation Safety Department for the constant technical assistance.

\end{document}